\documentclass[pra,twocolumn,groupedaddress,showpacs,floatfix]{revtex4}

\usepackage[utf8]{inputenc}
\usepackage{amsfonts}
\usepackage{amssymb}
\usepackage{boxedminipage}
\usepackage{listings}
\usepackage{minitoc}
\usepackage{ifpdf}

\ifpdf
\usepackage[pdftex]{graphicx}
\else
\usepackage{graphicx}
\fi

\newcommand{\ket}[1]{|#1 \rangle}
\newcommand{\bra}[1]{\langle #1|}

\newcommand{\ud}{\!\mathrm{d}}

\begin{document}

\title{Generating single-mode behavior in fiber-coupled optical cavities}
\author{Jonathan Busch\footnote{Corresponding author: j.busch@leeds.ac.uk} and Almut Beige}
\address{The School of Physics and Astronomy, University of Leeds, Leeds LS2 9JT, United Kingdom}

\date{\today}

\begin{abstract}
We propose to turn two resonant distant cavities effectively into one by coupling them via an optical fiber which is  coated with two-level atoms [Franson {\em et al.}, Phys.~Rev.~A {\bf 70}, 062302 (2004)]. The purpose of the atoms is to destructively measure the evanescent electric field of the fiber on a time scale which is long compared to the time it takes a photon to travel from one cavity to the other. Moreover, the boundary conditions imposed by the setup should support a small range of standing waves inside the fiber, including {\em one} at the frequency of the cavities. In this way, the fiber provides an additional decay channel for one common cavity field mode but not for the other. If the corresponding decay rate is sufficiently large, this mode decouples effectively from the system dynamics. A single non-local resonator mode is created. 
\end{abstract}


\maketitle

\section{Introduction} \label{intro}

Recent progress in experiments with optical cavities has mainly been motivated by potential applications in quantum information processing. These applications often require the simultaneous trapping of at least two atomic qubits inside a single resonator field mode. It has been shown that the common coupling to a quantised mode can be used for the implementation of quantum gate operations \cite{Pellizzari,Beige00,zheng,pachos,you} and the controlled generation of entanglement \cite{Cabrillo2,Marr,Plenio2,Metz,Metz2}. However, the practical realisation of these schemes with current technologies is experimentally challenging. The main reason is that strong atom-cavity interactions require relatively small mode volumes and high quality mirrors; aims that are difficult to reconcile with the placement of several atoms or ions into the same cavity. 

\begin{figure*}[t]
\begin{minipage}{1.9\columnwidth}
\begin{center}
\resizebox{\columnwidth}{!}{\rotatebox{0}{\includegraphics{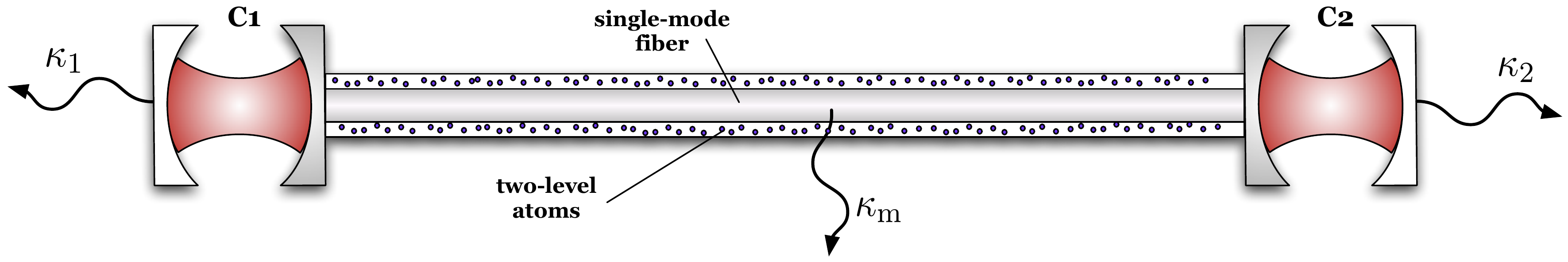}}}
\end{center}
\vspace*{-0.5cm} \caption{(Color online) Experimental setup of two optical cavities coupled via a single-mode fiber. Photons can leak out through the outer mirrors with the spontaneous decay rate $\kappa_1$ and $\kappa_2$, respectively. The connection between both cavities constitutes a third reservoir with spontaneous decay rate $\kappa_{\rm m}$ for a common non-local resonator field mode.} \label{scheme}
\end{minipage}
\end{figure*}

To solve this problem, it has been proposed to couple distant cavities via linear optics networks \cite{Cabrillo,Lim,Lim2}. Under realistic conditions, this strategy allows at least for the probabilistic build up of highly entangled states. Alternatively, one could shuttle atoms successively in and out of the resonator \cite{shuttling1,shuttling2,shuttling3}. In this paper we propose to use instead fiber-coupled cavities which employ reservoir engineering and similar ideas as in Ref.~\cite{BuschPRL} to turn two distant cavities effectively into one. Our aim is that atomic qubits placed into different cavities behave as if they were placed into the same cavity. When this becomes possible, quantum computing schemes designed for several qubits placed into the same resonator can be applied to a much wider range of experimental scenarios. They can be implemented with atomic qubits, quantum dots \cite{qdots,qdots2}, NV color centers \cite{NVC,NVC2}, and superconducting flux qubits \cite{Mooij}. Another possible application of fiber-coupled cavities is the transfer of information from one cavity to another \cite{Cirac,Pellizzari2,vanEnk}.

The experimental setup considered in this paper (c.f.~Fig.~\ref{scheme}) consists of two cavities with the same frequency $\omega_{\rm cav}$. Given two cavities with fixed polarization, there are two quantised cavity field modes. For example, one could describe the setup using the individual cavity modes with annihilation operators $c_1$ and $c_2$. But there is also the possibility of describing the cavities by two common (i.e.~non-local) field modes. Their cavity photon annihilation operators are of the general form
\begin{eqnarray} \label{com_modes}
c_a &=& \frac{1}{\xi} \left(\xi_2^* \, c_1 - \xi_1^* \, c_2 \right) \, , \nonumber \\
c_b &=& \frac{1}{\xi}(\xi_1 \, c_1 + \xi_2 \, c_2) \, ,
\end{eqnarray}
where $\xi_1$ and $\xi_2$ are complex coefficients and 
\begin{eqnarray} \label{xiii}
\xi &=& \sqrt{|\xi_1|^2 + |\xi_2|^2} \, .
\end{eqnarray}
One can easily check that, if $c_1$ and $c_2$ obey the usual boson commutator relations, then so do $c_a$ and $c_b$, 
\begin{eqnarray}
[ \, c_a,c_a^\dagger \, ] = [ \, c_b,c_b^\dagger \, ] &=& 1 ~~ {\rm and} ~~
[ \, c_a,c_b^\dagger \, ] = 0 \, .
\end{eqnarray}
An atomic qubit placed into one of the two cavities interacts in general with the $c_a$ and with the $c_b$ mode, since both are non-local. 

The purpose of the cavity fiber coupling with atomic coating shown in Fig.~\ref{scheme} is to asign different spontaneous decay rates to the $c_a$ and to the $c_b$ mode. If one of the two common cavity modes has a much larger spontaneous decay rate than the other one, it effectively decouples from the system dynamics \cite{BuschPRL,tonyfest}. A single non-local resonator mode is created. This means, atomic qubits placed into different cavities would indeed behave as if they were placed into the same cavity. 

To achieve this task, we impose the following conditions on the experimental setup considered in this paper:
\begin{enumerate}
\item  Different from Refs.~\cite{Bose,Bose2}, we do {\em not} treat the fiber as a resonant cavity with a single well-defined frequency. Instead, we assume boundary conditions which allow for a continuous range of frequencies which should include the cavity frequency $\omega_{\rm cav}$. This broadening of the fiber spectrum is in general due to the finite width of the fiber, imperfection of the mirrors, and the presence of atoms in its evanescent field \cite{Welsch}. 

\item At the same time, the frequency range supported by the fiber should not be too broad. The fiber needs to be short and thin enough to have a well defined optical path length for each frequency supported by the fiber. At the optical frequency $\omega_{\rm cav}$, there should be only {\em one} standing wave which fulfils the boundary condition of vanishing electric field amplitudes at the surface of the adjacent cavity mirrors. Standing waves which are half a wave length $\lambda_{\rm cav}$ shorter or longer should not fit into the fiber.

\item The single-mode fiber connecting the two cavities in Fig.~\ref{scheme} should be coated with two-level atoms. The purpose of the atoms is similar to their purpose in Ref.~\cite{Franson} by Franson {\em et al.}, namely to measure the evanescent electric field of the fiber and to provide an additional reservoir for the cavity photons. In the following, we assume that the atoms have a transition frequency $\omega_0$ and a non-zero decay rate $\Gamma$ such that they absorb light traveling through the fiber and dispose of it via spontaneous emission. In the following we denote the spontaneous decay rate associated with the leakage of photons out of the fiber by $\kappa_{\rm m}$.

\item The atoms should measure electric field amplitudes on a time scale which is long compared to the time it takes a photon to travel from one cavity to the other. In this way, the atoms measure only relatively long living photons inside the fiber, i.e.~the field amplitudes of the electromagnetic standing waves with vanishing amplitudes at the fiber ends. They should not be able to gain information about the source of a photon. 

\item Here we are especially interested in the parameter regime, where $\kappa_{\rm m}$ is much larger than the spontaneous decay rates $\kappa_1$ and $\kappa_2$ which describe the absorption of photons in the cavity mirrors and the leakage of photons into adjacent reservoirs other than the fiber. Moreover, $\kappa_{\rm m}$ should be much larger than any other coupling constants, like the Rabi frequencies $\Omega_1$ and $\Omega_2$ of externally applied laser fields, i.e.
\begin{eqnarray} \label{condi}
\kappa_{\rm m} &\gg & \kappa_i , \, \Omega_i \, .
\end{eqnarray}
In other words, cavity photons which leak out through the fiber decay on a much shorter time scale than the cavity photons which do not see this reservoir.
\end{enumerate}

Suppose, there is initially one photon in cavity 1 and none in cavity 2. In this case, some light will travel from cavity 1 to cavity 2. Once there is excitation in both cavities, the photons which do not couple to the one mode supported by the fiber at frequency $\omega_{\rm cav}$ can no longer enter the fiber. Other cavity photons leak more easily into the fiber, since their efforts are met by waves with the same amplitude coming from the other side. The above conditions assure that the photons are measured on a relatively slow time scale and that the atoms in the evanescent field of the fiber cannot distinguish photons traveling left or right. They can only absorb light which can exist for a relatively long time inside the fiber. This means, they only absorb photons from one common cavity mode but not from the other. There is hence a finite probability that the initial photon leaks relatively quickly into the environment. In addition, there is the possibility that the initial photon remains inside the setup for a relatively long time and becomes a shared photon between both cavities with no amplitude in the fiber. In the following we associate the cavity mode which does {\em not} see the fiber with the $c_a$ mode. The spontaneous decay rate $\kappa_a$ of photons in the $c_a$ mode is hence of a similar size as $\kappa_1$ and $\kappa_2$. The $c_b$ mode however sees the fiber reservoir in the middle in addition and has a decay rate comparable to $\kappa_{\rm m}$. Eq.~(\ref{condi}) hence implies
\begin{eqnarray}
\kappa_b &\gg & \kappa_a , \, \Omega_i  
\end{eqnarray}
which is exactly what we want to achieve.

\begin{figure}[t]
\begin{minipage}{\columnwidth}
\begin{center}
\resizebox{\columnwidth}{!}{\rotatebox{0}{\includegraphics{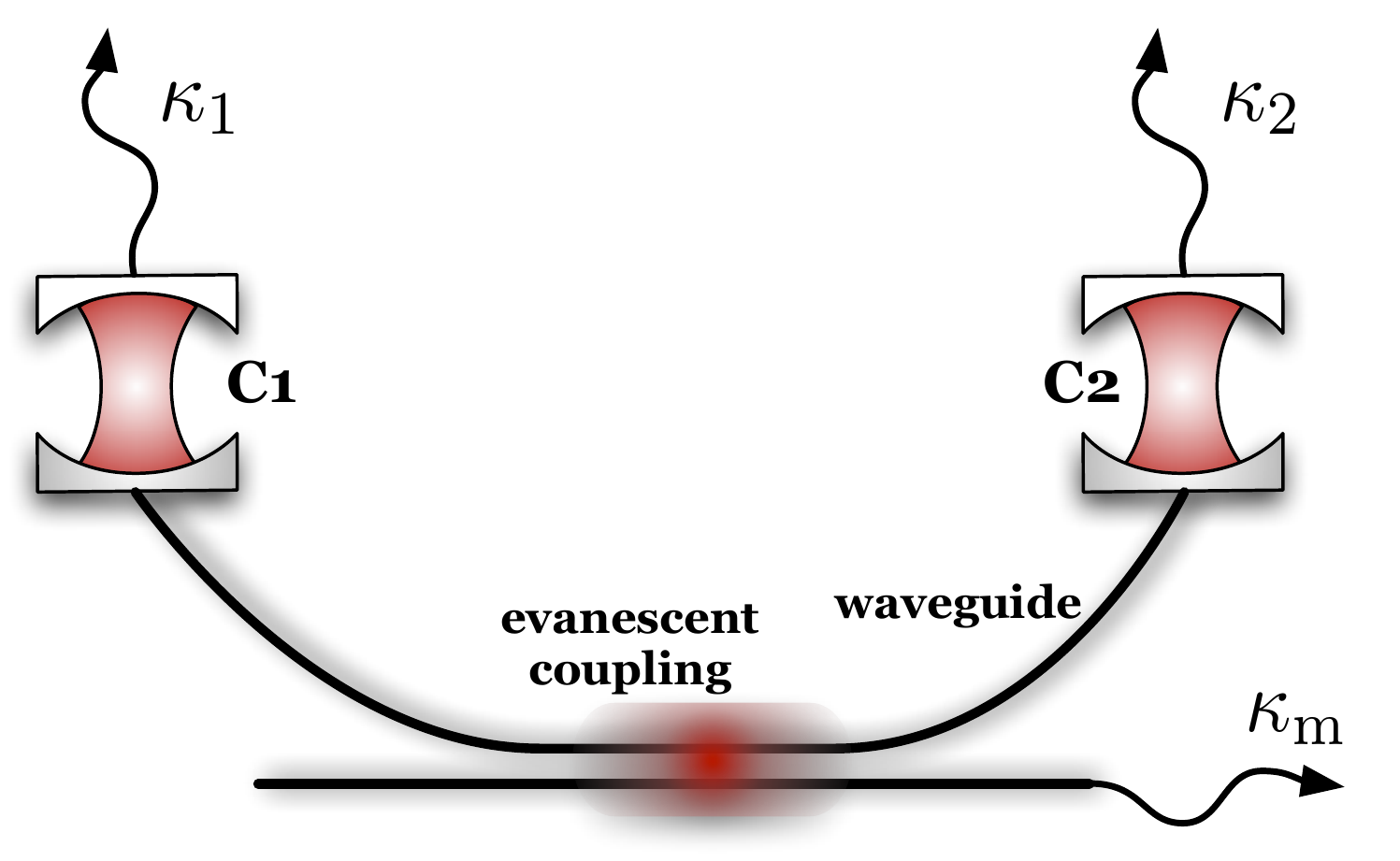}}}
\end{center}
\vspace*{-0.5cm} \caption{(Color online) Schematic view of an alternative experimental setup. If the cavities are mounted on an atom chip, the could be coupled via a waveguide etched onto the chip. To emulate environment-induced measurements of the field amplitude within the waveguide, a second waveguide should be placed into its evanescent field which constantly damps away any eletromagnetic field amplitudes.} \label{scheme2}
\end{minipage}
\end{figure}

Although this paper refers in the following only to single-mode fibers, any coupling of the cavities which meets the above requirements would work equally well. One possible alternative is shown in Fig.~\ref{scheme2}. If the cavities are mounted on an atom chip, a similar connection between them could be created with the help of a waveguide (or nanowire) etched onto the chip. Such a connection too supports only a single electromagnetic field mode. To detect its field amplitude, a second waveguide connected to a detector should be placed into its evanescent field, thereby constantly removing any field amplitude from the waveguide between the cavities.

Fiber coupled optical cavities with applications in quantum information processing have already been widely discussed in the literature (see e.g.~Refs.~\cite{Bose,Bose2,Pellizzari2,vanEnk,Cirac,Parkins}). The main difference of the cavity coupling scheme presented here is that it does not rely on coherent time evolution. Instead it actively uses dissipation in order to achieve its task. We therefore expect that the proposed scheme is more robust against errors. For example, the fiber considered here which is coated with two-level atoms acts as a reservoir for the cavity photons and supports a continuous range of frequencies. It can hence be longer than when it needs to contain only a single frequency as in Refs.~\cite{Bose,Bose2,Pellizzari2,vanEnk}. In addition, the setup considered here is robust against fiber losses \cite{Cirac,Parkins}.

An alternative scheme for the generation of single-mode behavior in distant optical cavities has recently been proposed by us in Ref.~\cite{BuschPRL}. Different from the setup in Fig.~\ref{scheme}, we considered two optical cavities with each of them individually coupled to an optical single-mode fiber. These fibers guide photons from each cavity onto a single photon detector which cannot resolve the origin of the incoming photons. Despite its similarity with the two-atom double slit experiment by Eichmann {\em et al.} \cite{Eichmann,Schoen,pachos00}, a Gaussian beam analysis of the scheme proposed in Ref.~\cite{BuschPRL} shows that achieving real indistinguishability would require optical fibers with a diameter much smaller than the optical wavelength \cite{Busch}. These are relatively hard to realise experimentally although this is feasible with current technology \cite{Arno,Jim}. The setup in Fig.~\ref{scheme} avoids the use of subwavelength fibers by replacing them with a  single {\em naturally-aligned} fiber. 

\begin{figure}[t]
\begin{minipage}{\columnwidth}
\begin{center}
\resizebox{\columnwidth}{!}{\rotatebox{0}{\includegraphics{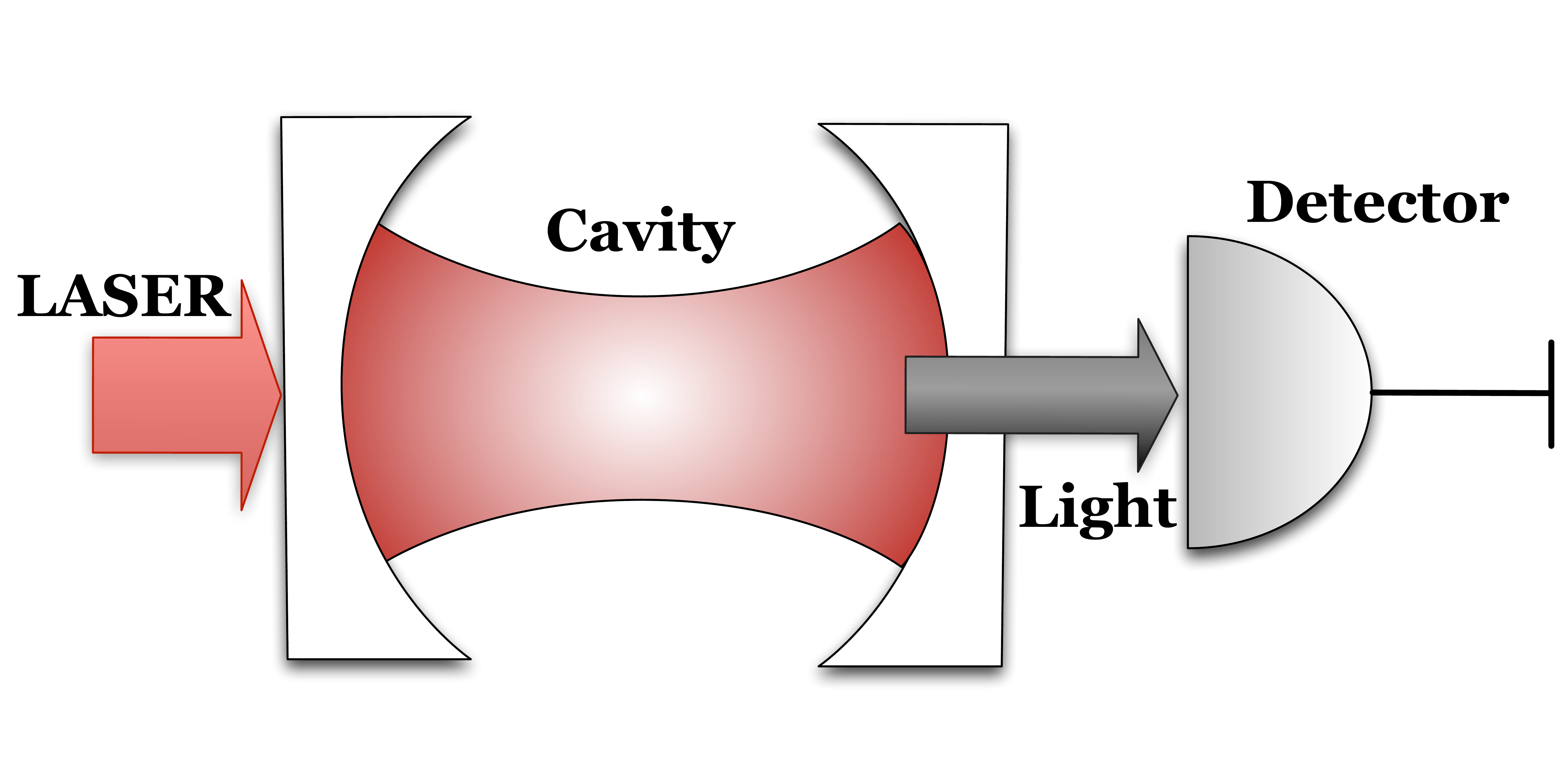}}}
\end{center}
\vspace*{-0.5cm} \caption{(Color online) Experimental setup of a single cavity driven by a laser field. The photons leaking out through the cavity mirrors are monitored by a detector.} \label{cavitypic}
\end{minipage}
\end{figure}

There are five sections in this paper. Section \ref{single_cav} gives an overview over the open system description of a single cavity, thereby providing a blueprint for the expected behavior of the setup in Fig.~\ref{scheme}. Section \ref{two_cavs} includes a detailed derivation of the master equation for two fiber-coupled cavities. Section \ref{opsmode} describes two different scenarios for which the $c_b$ mode decouples effectively from the system dynamics with one of them being especially robust against parameter fluctuations. Finally, we summarise our findings in Section \ref{conc}.

\section{Open system approach for a single laser-driven cavity} \label{single_cav}

In this section we describe how to predict the possible quantum trajectories of an optical cavity which is driven by a resonant laser field and continuously leaks photons through its cavity mirrors, as shown in Fig.~\ref{cavitypic}. We derive the master equation for this setup by adopting the quantum jump approach introduced in Refs.~\cite{Hegerfeldt,Molmer,Carmichael} and calculate its stationary state photon emission rate. Later we refer to the equations in this section when deriving the master equation for two fiber-coupled optical cavities and when discussing conditions for their single-mode behaviour. 

\subsection{System Hamiltonian}

The setup in Fig.~\ref{cavitypic} consists of an optical cavity which interacts with the surrounding free radiation field and is driven by a resonant laser field. Its Hamiltonian is hence of the form
\begin{eqnarray} \label{HHH}
	H = H_{\rm cav} + H_{\rm res} + H_{\rm dip} \, ,
\end{eqnarray}
where $H_{\rm cav}$ is the cavity Hamiltonian, $H_{\rm res}$ is the reservoir Hamiltonian, and $H_{\rm dip}$ takes the dipole coupling of the cavity to the driving laser field and the environment into account. If we denote the frequency of the cavity mode and the modes of the free radiation field with wave number $k$ by $\omega_{\rm cav}$ and $\omega_k$ and the corresponding photon annihilation operators by $c$ and $a_k$, respectively, then
\begin{eqnarray} \label{Hcav-res}
	H_{\rm cav} &=& \hbar\omega_{\rm cav} \, c^{\dagger} c \, , \nonumber \\
	H_{\rm res} &=&  \sum_k \hbar \omega_k \, a_k^{\dagger}a_k \, .
\end{eqnarray}
Here we assume that the polarisation of the applied laser field, the cavity field, and the modes of the free radiation field is the same. As long as no mixing of different polarisation modes occurs, these are the only modes which have to be taken into account. Moreover, we have
\begin{eqnarray} \label{Hdip}
	H_{\rm dip} = e{\bf D}\cdot ({\bf E}_{\rm laser} (t) + {\bf E}_{\rm res}) \, ,
\end{eqnarray}
where $\mathbf{D} \propto c + c^{\dagger}$ is the effective dipole moment of the cavity mode and where ${\bf E}_{\rm laser}(t)$ and ${\bf E}_{\rm res}$ are the electric fields of the driving laser and of the free radiation field, respectively. 

Treating the laser field with frequency $\omega_{\rm L} = \omega_{\rm cav}$ as a classical field whilst considering the modes of the reservoir quantised, this Hamiltonian can be written as 
\begin{eqnarray} \label{Hdip2}
	H_{\rm dip} &=& \frac{1}{2} \hbar \Omega \, {\rm e}^{i \omega_{\rm cav} t} \, c + \sum_k \hbar g_k \, c a_k^{\dagger} + {\rm H.c.} \, , 
\end{eqnarray}
where the rotating wave approximation has already been applied. Here $\Omega$ is the (complex) laser Rabi frequency and the $g_k$ are the (complex) coupling constants of the interaction between the cavity and the free radiation field due to overlapping electric field modes in the vicinity of the resonator mirrors. 

For simplicity, we now move into the interaction picture with respect to the interaction-free Hamiltonian 
\begin{eqnarray} \label{H0}
H_0 = H_{\rm cav} + H_{\rm res} \, .
\end{eqnarray}
In this case, the Hamiltonian of system and environment simplifies to 
\begin{eqnarray} \label{HI}
	H_{\rm I} &=& \frac{1}{2} \hbar \Omega \, c + \sum_k \hbar g_k \, {\rm e}^{i (\omega_k - \omega_{\rm cav}) t } \, c a_k^{\dagger} + {\rm H.c.}
\end{eqnarray}
The laser field simply creates and annihilates photons inside the cavity mode, while the cavity-reservoir coupling results in an exchange of photon energy between system and environment. 

\subsection{No-photon time evolution}

As in Refs.~\cite{Hegerfeldt,Molmer,Carmichael} we assume in the following that the environment constantly performs measurements on the free radiation field whether a photon has been emitted or not. In quantum optical systems, there are in general no photons in the free radiation field, since these travel away (or are absorbed by the environment) such that they cannot return into the system. In the following, we assume therefore that the cavity is initially in a state $|\varphi(0) \rangle$, while the free radiation field is in its vacuum state $|0_{\rm ph} \rangle$. Using the projection postulate for ideal measurements, one can then show that the state of the system equals 
\begin{eqnarray}\label{proj}
\ket{0_{\rm ph}} |\varphi_0 (\Delta t) \rangle &=& \ket{0_{\rm ph}} \bra{0_{\rm ph}}U_{\rm I}(\Delta t,0)\ket{0_{\rm ph}} |\varphi(0) \rangle ~~~~
\end{eqnarray}
at $\Delta t$ under the condition of no photon emission. A comparison of both sides of this equation shows that 
\begin{eqnarray}
|\varphi_0 (\Delta t) \rangle = U_{\rm cond}(\Delta t,0) |\varphi(0) \rangle
\end{eqnarray}
with the conditional time evolution operator defined as
\begin{eqnarray}\label{Ucond_def}
	U_{\mathrm{cond}}(\Delta t,0) \equiv \bra{0_{\rm ph}}U(\Delta t,0)\ket{0_{\rm ph}} \, .
\end{eqnarray}
The probability for no photon emission in $\Delta t$ can now be written as $\| U_{\rm cond}(\Delta t,0) |\varphi(0) \rangle \|^2$.

Using Eq.~(\ref{HI}) and second order perturbation theory, and proceeding as in Refs.~\cite{Hegerfeldt,Molmer,Carmichael}, one can easily show that the corresponding conditional time evolution operator equals
\begin{eqnarray} \label{U_cond}
U_{\mathrm{cond}}(\Delta t,0) 
	&=& \mathbb{I} - \frac{\rm i}{2} \, \big( \Omega \, c + \Omega^* \, c^{\dagger} \big) \Delta t \nonumber \\
	&& \hspace*{-1.7cm} - \sum_k g_k^2 \int_0^{\Delta t} \ud t \int_0^t \ud t' \, {\rm e}^{{\rm i}(\omega_k - \omega_{\rm cav})(t' - t)} \, c^\dagger c \, . ~~~
\end{eqnarray}
To evaluate the double integral in this equation, we substitute $t'$ by $\tau \equiv t - t'$. Considering a time interval $\Delta t$ with $\Delta t \gg 1/\omega_{\rm cav}$, the second integral can be replaced by a $\delta$-function. Neglecting a term corresponding to a level shift which can be absorbed into $H_{\rm cav}$ of the total system Hamiltonian, we obtain
\begin{eqnarray} \label{deltafunc}
	\int_0^{\Delta t} \ud t \int_0^t \ud t' \, {\rm e}^{{\rm i}(\omega_k - \omega_{\rm cav})(t' - t)} = 
	\pi \delta(\omega_k - \omega_{\rm cav}) \Delta t \, .
\end{eqnarray}
The conditional Hamiltonian corresponding to the time evolution in Eq.~(\ref{U_cond}) hence equals 
\begin{eqnarray} \label{H_cond}
	H_{\mathrm{cond}} = \frac{1}{2}\hbar \Omega \, c + {\rm H.c.} - \frac{\rm i}{2} \hbar\kappa \, c^{\dagger}c 
\end{eqnarray}
with the spontaneous cavity leakage rate $\kappa$. Suppose we denote the cavity-environment coupling constant $g_k$ for the mode which is resonant with the cavity field by $g_{\rm c}$. Then $\kappa $ can be written as
\begin{eqnarray} \label{calN}
	\kappa = {2 \pi \over {\cal N}} \, g_{\rm c} ^2 \, , 
\end{eqnarray}
where ${\cal N}$ is a normalisation factor which depends for example on the quantisation volume of the reservoir. The second term in Eq.~(\ref{H_cond}) takes into account that not seeing a photon gradually reveals information about the system, thereby increasing the relative population in states with lower photon numbers.

\subsection{Effect of photon emission} \label{photonemi}

Analogously, one can derive the state of the system in case of an emission which we write in the following as 
\begin{eqnarray}
|\varphi_{\rm ph} (\Delta t) \rangle &=& R \, |\varphi(0) \rangle \, .
\end{eqnarray}
Replacing the no-photon projector $|0_{\rm ph} \rangle \langle 0_{\rm ph}|$ in Eq.~(\ref{proj}) by the projector onto all states with at least one photon in the free radiation field, using first order perturbation theory, and proceeding again as in Refs.~\cite{Hegerfeldt,Molmer,Carmichael}, we find that $R$ equals
\begin{eqnarray} \label{Rsingle}
	R = \sqrt{\kappa} \, c  \, .
\end{eqnarray}
Here the normalisation of the reset operator $R$ has been chosen such that $\| R \, |\varphi(0) \rangle \|^2 \, \Delta t$ is the probability for a photon emission in $\Delta t$.

\subsection{Master equation} \label{sec:master}

Averaging over both possibilities, i.e.~over a subensemble of cavities without and a subensemble of cavities with photon emission in $\Delta t$, we move from the above described quantum jump approach \cite{Hegerfeldt,Molmer,Carmichael} to the master equation. Doing so, we find that the density matrix of the cavity field evolves according to 
\begin{eqnarray} \label{master}
	\dot{\rho} = -\frac{\rm i}{\hbar} \, \big[\, H_{\rm cond}\rho - \rho H_{\rm cond}^{\dagger} \, ] + R \, \rho \, R^{\dagger} \, .
\end{eqnarray}
This is the standard master equation for the quantum optical description of the field inside an optical cavity. 

\subsection{Stationary state photon emission rate} \label{sec:fluorescence}

If we are for example interested in the time evolution of the mean number of photons $n$ inside the cavity, then there is no need to solve the whole master equation (\ref{master}). Instead, we use this equation to get a closed set of rate equations with $n$ being one of its variables. More concretely, considering the expectation values 
\begin{eqnarray}
	n &\equiv & \langle c^{\dagger}c \rangle \, , \nonumber \\
	k &\equiv  & \frac{\rm i}{|\Omega|} \langle \Omega \, c - \Omega^* \, c^{\dagger} \rangle \, ,
\end{eqnarray}
we find that their time evolution is given by 
\begin{eqnarray}
	\dot n &=& \frac{1}{2}|\Omega| \, k  - \kappa \, n \, , \nonumber \\
	\dot k &=& |\Omega| - \frac{1}{2}\kappa \, k \, . 
\end{eqnarray}
Setting the right hand sides of these equations equal to zero, we find that the stationary state of the laser-driven cavity corresponds to $n = |\Omega|^2 /\kappa^2$. Since the steady state photon emission rate is the product of $n$ with the decay rate $\kappa$, this yields
\begin{eqnarray} \label{III}
	I &=& |\Omega|^2 / \kappa \, .
\end{eqnarray}
Measurements of the parameter dependence of this intensity can be used to determine $|\Omega |$ and $\kappa$ experimentally. 

\section{Open system approach for two laser-driven fiber-coupled cavities} \label{two_cavs}

In this section, we derive the master equation for the two fiber-coupled optical cavities shown in Fig.~\ref{scheme}. We proceed as in the previous section and obtain their master equation by averaging again over a subensemble with and a subsensemble without photon emission. A discussion of the behavior predicted by this equation for certain interesting parameter regimes can be found later in Section \ref{opsmode}. 

\subsection{System Hamiltonian}

The total system Hamiltonian $H$ for the setup in Fig.~\ref{scheme} in the Schr\"odinger picture is of exactly the same form as the Hamiltonian in Eq.~(\ref{HHH}). Again, $H_{\rm cav}$ and $H_{\rm res}$ denote the energy of the system and its reservoirs, while $H_{\rm dip}$ models the cavity-environment couplings and the effect of applied laser fields. In the following, we denote the annihilation operators of the two cavities by $c_1$ and $c_2$, respectively, while 
\begin{eqnarray}
\omega_{\rm c,1} \, = \, \omega_{\rm c,2} \, = \, \omega_{\rm cav}
\end{eqnarray}
is the corresponding frequency which should be for both cavities the same. In analogy to Eq.~(\ref{Hcav-res}), the energy of the resonators is hence given by  
\begin{eqnarray}
	H_{\rm cav} &=& \sum_{i=1,2}  \hbar \omega_{\rm cav} \, c_i^{\dagger} c_i \, .
\end{eqnarray}
The reservoir of the system now consists of three components. Its Hamiltonian can be written as
\begin{eqnarray}
	H_{\rm res} = \sum_{i=1,2}\sum_k  \hbar  \omega_k \, a_{k,i}^{\dagger} a_{k,i} + \sum_k \hbar  \omega_k \, b_k^{\dagger} b_k \, ,
\end{eqnarray} 
where $\omega_k$ denotes the frequency of the free field radiation modes with wavenumber $k$. The annihilation operators $a_{k,i}$ describe the free radiation field modes on the unconnected side of each cavity with $k$ being the respective wavenumber and $i$ indicating which cavity the field interacts with. The annihilation operators $b_k$ describe the continuum of quantised light modes in the optical single-mode fiber with vanishing electric field amplitudes at the fiber ends. For each wave number $k$, these modes correspond to a single standing light wave with contributions traveling in different directions through the fiber. As in the previous section, we restrict ourselves to the polarisation of the applied laser field. Since there is no polarisation mode mixing, this is the only polarisation which needs to be taken into account.  

The only term still missing is the interaction Hamiltonian $H_{\rm dip}$ which describes the coupling of the two cavities to their respective laser fields and to their respective reservoirs. Assuming that both lasers in Fig.~\ref{scheme} are in resonance and applying the usual dipole and rotating wave approximation, $H_{\rm dip}$ can in analogy to Eq.~(4) in Ref.~\cite{Pellizzari2}, be written as 
\begin{eqnarray}
	H_{\rm dip} &=& \sum_{i=1,2}  \sum_k \hbar s_{k,i} \, c_i a_{k,i}^\dagger 
	+ \hbar g_{k,i} \, c_i b_k^\dagger \nonumber \\
	&& + \sum_{i=1,2} \frac{1}{2} \hbar \Omega_i \, {\rm e}^{- {\rm i} \omega_{\rm cav} t} \, c_i + {\rm H.c.} \, , ~~~
\end{eqnarray}
where $s_{k,i}$ and $g_{k,i}$ are system-reservoir coupling constants and where $\Omega_i$ is the Rabi frequency of the laser driving cavity $i$. 

To calculate the photon and the no-photon time evolution of the cavities over a time interval $\Delta t$ with the help of second order perturbation theory, we proceed as in Section \ref{single_cav} and transform the Hamiltonian $H$ of the system into the interaction picture relative to $H_0$ in Eq.~(\ref{H0}). This finally yields 
\begin{eqnarray} \label{HIHI}
	H_{\rm I} &=& \sum_{i=1,2} \sum_k \hbar s_{k,i} \, {\rm e}^{{\rm i}(\omega_k - \omega_{\rm cav})t} \, c_i a_{k,i}^{\dagger} \nonumber \\
	&& + \hbar g_{k,i} \, {\rm e}^{{\rm i}(\omega_k - \omega_{\rm cav})t} c_i b_k^\dagger + \frac{1}{2} \hbar \Omega_i \, c_i  + {\rm H.c.} ~~
\end{eqnarray}
which describes the interaction of the cavities with their reservoirs and the two lasers.

\subsection{No-photon time evolution}

As in the previous section, in the single-cavity case, we assume that the unconnected mirrors of the resonators leak photons into free radiation fields, where they are continuously monitored by the environment or actual detectors. In addition, there is now a continuous monitoring of the photons which can leak into the single-mode fiber connecting both cavities.  Again, it is not crucial whether an external observer actually detects these photons or not, as long as the effect on the system is the same as if the photon has actually been measured. Important is only that photons within the three reservoirs, the surrounding free radiation fields and the single-mode fiber, are constantly removed from the system and cannot re-enter the cavities. 

In principle, there are now three different response times $\Delta t$ of the environment, i.e.~one for each reservoir. For simplicity and since it does not affect the resulting master equation we consider only one of them. Denoting this response time of the environment again by $\Delta t$, we assume in the following that
\begin{eqnarray} \label{anti}
	\frac{1}{\omega_{\rm cav}} \ll \Delta t ~~ {\rm and} ~~ \Delta t \ll \frac{1}{\kappa_{\rm m}} , \, \frac{1}{\kappa_1}, \, \frac{1}{\kappa_2} \, ,
\end{eqnarray}
where $\kappa_{\rm m}$ is the spontaneous decay rate for the leakage of photons from the cavities into the optical fiber, while $\kappa_i$ denotes the decay rate of cavity $i$ with respect to its outcoupling mirror. The conditions in Eq.~(\ref{anti}) allow us to calculate the time evolution of the system within $\Delta t$ with second order perturbation theory. The first condition assures that there is sufficient time between measurements for photon population to build up within the reservoirs \footnote{Otherwise, there would not be any spontaneous emissions.}. The second condition avoids the return of photons from the reservoirs into the cavities. 

Proceeding as in the previous section and using again Eq.~(\ref{Ucond_def}), we find that the conditional Hamiltonian describing the time evolution of the two cavities under the condition of no photon emission in $\Delta t$ into any of the three reservoirs equals 
\begin{eqnarray}\label{U_cond-2}
	&& \hspace*{-0.5cm} U_{\rm cond}(\Delta t,0) = \nonumber \\
	&&\mathbb{I} - \frac{\rm i}{2} \, \sum_{i=1,2} \big( \Omega_i \, c_i + \Omega_i^* \, c_i^{\dagger} \big) \Delta t \nonumber \\
	&&- \int_0^{\Delta t} \ud t \int_0^{t} \ud t' \sum_{i=1,2} \sum_k {\rm e}^{{\rm i}(\omega_k - \omega_{\rm cav})(t'-t)} |s_{k,i}|^2 \, c_i^\dagger c_i \nonumber \\
	&&- \int_0^{\Delta t}\ud t \int_0^{t} \ud t' \sum_k {\rm e}^{{\rm i}(\omega_k - \omega_{\rm cav})(t'-t)} ( g_{k,1}^* \, c_1^{\dagger} + g_{k,2}^* \, c_2^{\dagger}) \nonumber \\
	&& \qquad \qquad \times(g_{k,1} \, c_1 + g_{k,2} \, c_2) \, .
\end{eqnarray}
In analogy to Eq.~(\ref{H_cond}), the first three terms evaluate to 
\begin{eqnarray}
	&&\mathbb{I} - \frac{\rm i}{2} \, \sum_{i=1,2} \big( \Omega_i \, c_i + \Omega_i^* \, c_i^{\dagger} \big) \Delta t \nonumber \\
	&&- \frac{1}{2} \kappa_1\Delta t \, c_1^{\dagger}c_1 - \frac{1}{2} \kappa_2 \Delta t \, c_2^{\dagger}c_2 \, .
\end{eqnarray}
Using exactly the same approximations as in the previous section and the notation
\begin{eqnarray}
\xi_i \equiv \sum_{k} g_{k,i} \, .
\end{eqnarray}
With $\xi$ defined as in Eq.~(\ref{xiii}), the final term in Eq.~(\ref{U_cond-2}) can be written as 
\begin{eqnarray}
	-\frac{1}{2 \xi^2} \kappa_{\rm m} \Delta t \, ( \xi_1^* \, c_1^{\dagger} + \xi_2^* \, c_2^{\dagger})\left( \xi_1 \, c_1 +  \xi_2 \, c_2 \right) \, .
\end{eqnarray}
Here $\kappa_1$, $\kappa_2$, and $\kappa_{\rm m}$ are the spontaneous decay rates already mentioned in Eq.~(\ref{anti}). The corresponding conditional Hamiltonian equals
\begin{eqnarray}\label{H_cond-mix}
	H_{\rm cond} &=& \sum_{i=1,2} \frac{1}{2} \hbar \Omega_i \, c_i  + {\rm H.c.} - \frac{{\rm i}}{2} \hbar \kappa_i \, c_i^{\dagger}c_i  \nonumber \\
		&& - \frac{{\rm i}}{2 \xi^2} \hbar \kappa_{\rm m} \, (\xi_1^* \, c_1^{\dagger} + \xi_1^* \, c_2^{\dagger} ) (\xi_1 \, c_1 + \xi_2 \, c_2 ) ~~~
\end{eqnarray}
and describes the no-photon time evolution of cavity 1 and cavity 2.

\subsection{Effect of photon emission}

Proceeding as in Section~\ref{photonemi}, assuming that the respective reservoir is initially in its vacuum state, using first order perturbation theory, and calculating the state of the system under the condition of a photon detection, we find that photon emission into the individual reservoir of cavity $i$ is described by 
\begin{eqnarray}\label{reset2}
	R_i &=& \sqrt{\kappa_i} \, c_i \, . 
\end{eqnarray}
The leakage of a photon through the fiber reservoir changes the system according to
\begin{eqnarray}\label{reset22}
      R_{\rm m} &=& {1 \over \xi} \sqrt{\kappa_{\rm m}} \, (\xi_1 \, c_1 + \xi_2 \, c_2) \, .
\end{eqnarray}
The normalisation of these operators has again been chosen such that the probability for an emission in $\Delta t$ into one of the reservoirs equals $\| R_{\rm x} \, |\varphi(0) \, \rangle \|^2 \, \Delta t$ with ${\rm x}=1,2,{\rm m}$ and with  $|\varphi (0) \rangle$ being the initial state of the two cavities. 

\subsection{Master equation}

Averaging again over the possibilities of both no-photon evolution and photon emission events, we arrive at the master equation
\begin{eqnarray}\label{master2}
	\dot{\rho} &=& - {{\rm i} \over \hbar} \left[ \, H_{\rm cond} , \rho \, \right] + R_1 \, \rho \, R_1^\dagger + R_2 \, \rho \, R_2^\dagger  \nonumber \\
	&& + R_{\rm m} \, \rho \, R_{\rm m}^\dagger
\end{eqnarray}
which is analogous to Eq.~(\ref{master}) but where $\rho$ is now the density matrix of the two cavity fields.

\section{Single-mode behavior of two fiber-coupled cavities} \label{opsmode}

In this section, we discuss how to decouple one of the common cavity field modes in Eq.~(\ref{com_modes}) from the system dynamics \cite{tonyfest}. After introducing a certain convenient common mode representation, we see that there are two interesting parameter regimes: The first one is defined by a careful alignment of the Rabi frequencies $\Omega_1$ and $\Omega_2$, whilst the second one is defined by the condition that $\kappa_{\rm m}$ is much larger than all other spontaneous decay rates and laser Rabi frequencies in the system, as assumed in Eq.~(\ref{condi}). In this second parameter regime, one of the common modes can be adiabatically eliminated from the system dynamics. Consequently, this case does not require any alignment and is much more robust against parameter fluctuations. As we shall see below, the resulting master equation and its stationary state photon emission rate are formally the same as those obtained in Section \ref{sec:fluorescence} for the single-cavity case.

\subsection{Common mode representation}

Looking at the conditional Hamiltonian in Eq.~(\ref{H_cond-mix}), it is easy to see that $\kappa_{\rm m}$ is the spontaneous decay of a certain single non-local cavity field mode. Adopting the notation introduced in Section \ref{intro}, we see that this mode is indeed the $c_b$ mode defined in Eq.~(\ref{com_modes}). As already mentioned in the Introduction, the $c_b$ mode is the only common cavity field which interacts with the optical fiber connecting both cavities. The fiber provides an additional reservoir into which the photons in this mode can decay with $\kappa_{\rm m}$ being the corresponding spontaneous decay rate. Photons in the $c_a$ mode do not see the fiber and decay only via $\kappa_1$ and $\kappa_2$.

It is hence natural to replace the annihilation operators $c_1$ and $c_2$ by the common mode operators $c_a$ and $c_b$. Doing so, Eq.~(\ref{H_cond-mix}) becomes
\begin{eqnarray}\label{H_cond-comm}
H_{\rm cond} &=& \frac{1}{2} \hbar (\Omega_a \, c_a + \Omega_b \, c_b) + {\rm H.c.} -\frac{\rm i}{2} \hbar \kappa_{\rm m} \, c_b^{\dagger}c_b ~~ \nonumber \\
&& - \frac{\rm i}{2 \xi^2} \hbar \, \Big[ \left(  \kappa_1 |\xi_2|^2 +  \kappa_2 |\xi_1|^2 \right) c_a^\dagger c_a   \nonumber \\
&& + \left(  \kappa_1 |\xi_1|^2 +  \kappa_2 |\xi_2|^2 \right) c_b^\dagger c_b \nonumber \\
&& + \left(  \kappa_1 - \kappa_2 \right) \left( \xi_1 \xi_2 \, c_b^\dagger c_a + \xi_1^* \xi_2^* \, c_a^\dagger c_b \right) \Big] 
\end{eqnarray}
with the effective Rabi frequencies
\begin{eqnarray}\label{Omab}
	\Omega_a &\equiv&  \frac{1}{\xi}(\Omega_1 \xi_2 - \Omega_2 \xi_1) \, , \nonumber \\
	\Omega_b &\equiv&  \frac{1}{\xi}(\Omega_1 \xi_1^* + \Omega_2 \xi_2^*) \, .
\end{eqnarray}
The last term in Eq.~(\ref{H_cond-comm}) describes a mixing of the $c_a$ mode and the $c_b$ mode which occurs when the decay rates $\kappa_1$ and $\kappa_2$ are not of the same size. Finally, we find that the reset operators in Eqs.~(\ref{reset2}) and (\ref{reset22}) become
\begin{eqnarray}\label{R_common}
	R_1 &=& \frac{1}{\xi}  \sqrt{\kappa_1} \, (\xi_2 \, c_a + \xi_1^* \, c_b) \, , \nonumber \\
	R_2 &=& - \frac{1}{\xi} \sqrt{\kappa_2} \, ( \xi_1 \, c_a -  \xi_2^* \, c_b) \, , \nonumber \\
	R_{\rm m} &=& \sqrt{\kappa_{\rm m}} \, c_b 
\end{eqnarray}
in the common mode representation. 

\subsection{Single-mode behavior due to careful alignment}\label{align}

Let us first have a look at the case where the single-mode behavior of the two cavities in Fig.~\ref{scheme} is due to a careful alignment of the Rabi frequencies $\Omega_1$ and $\Omega_2$ and both cavity decay rates being the same, i.e. 
\begin{eqnarray} \label{44}
	\kappa \equiv \kappa_1 = \kappa_2
\end{eqnarray}
which sets $\kappa_1 - \kappa_2$ equal to zero. When two fiber-coupled cavities are driven by two laser fields with a fixed phase relation, the result is always the driving of only one common cavity field mode. If the cavities are therefore driven such that the driven mode is identical to the $c_a$ mode, an initially empty $c_b$ mode remains empty. As one can easily check using the definitions of the Rabi frequencies $\Omega_a$ and $\Omega_b$ in Eq.~(\ref{Omab}), this applies when
\begin{eqnarray}\label{alignOmega}
	{\Omega_1 \over \Omega_2} &=& - {\xi_2^* \over \xi_1^*} \, ,  
\end{eqnarray}
as it results in $\Omega_a \neq 0 $ and $\Omega_b = 0$. 

The question that now immediately arises is how to choose $\Omega_1$ and $\Omega_2$ in an experimental situation where $\xi_1$ and $\xi_2$ are not known. We therefore remark here that the sole driving of the $c_a$ mode can be distinguished easily from the sole driving of the $c_b$ mode by actually measuring the photon emission through the optical fiber. In the first case, the corresponding stationary state photon emission rate assumes its minimum, while it assumes its maximum in the latter. Variations of the Rabi frequency $\Omega_1$ with respect to $\Omega_2$ in a regime where both of them are of comparable size as $\kappa_{\rm m}$ can hence be used to determine $\xi_1/\xi_2$ experimentally.

Neglecting all terms which involve the annihilation operator $c_b$, as there are no $c_b$ modes to annihilate, results in the effective master equation
\begin{eqnarray}\label{master4}
	\dot{\rho} &=& - {{\rm i} \over \hbar} \left[ \, H_{\rm cond} , \rho \, \right] + \kappa \, c_a \, \rho \, c_a^\dagger \, , \nonumber \\
	H_{\rm cond} &=& \frac{1}{2} \hbar \Omega_a \, c_a + {\rm H.c.} - \frac{\rm i}{2} \hbar \, \kappa \, c_a^\dagger c_a \, .
\end{eqnarray}
This master equation is equivalent to Eqs.~(\ref{H_cond}), (\ref{Rsingle}), and (\ref{master}) in Section \ref{single_cav} which describes a single cavity. However, it is important to remember that the above equations are only valid when the alignment of the laser Rabi frequencies and cavity decay rates is {\it exactly} as in Eqs.~(\ref{44}) and (\ref{alignOmega}). Any fluctuation forces us to reintroduce the $c_b$ mode into the description of the system dynamics. 

\subsection{Robust decoupling of one common mode}

To overcome this problem, let us now have a closer look at the parameter regime in Eq.~(\ref{condi}), where the laser Rabi frequencies $\Omega_a$ and $\Omega_b$, and the spontaneous decay rates $\kappa_1$ and $\kappa_2$ are much smaller than $\kappa_{\rm m}$. To do so, we write the state vector of the system under the condition of no photon emission as
\begin{eqnarray} 
	|\varphi^0 (t) \rangle &= & \sum_{i,j=0}^\infty \zeta_{i,j}(t) \, |i,j \rangle \, ,
\end{eqnarray}
where $|i,j \rangle$ denotes a state with $i$ photons in the $c_a$ mode and $j$ photons in the $c_b$ mode and the 
$\zeta_{i,j}(t)$ are the corresponding coefficients of the state vector at time $t$. Using Eqs.~(\ref{master2}), (\ref{H_cond-comm}), and (\ref{R_common}) one can then show that the time evolution of the coefficients $\zeta_{i,0}$ and $\zeta_{i,1}$ is given by
\begin{eqnarray}\label{pop0}
	\dot{\zeta}_{i,0} &=& -\frac{\rm i}{2} \left[ \sqrt{i+1} \Omega_a \zeta_{i+1,0} + \sqrt{i} \Omega_a^*  \zeta_{i-1,0} + \Omega_b \zeta_{i,1} \right] \nonumber \\
	&&-\frac{1}{2\xi^2}\kappa_1 \left[ i |\xi_2|^2 \zeta_{i,0} + \sqrt{i} \xi_1^*\xi_2^* \zeta_{i-1,1} \right] \nonumber \\
	&&-\frac{1}{2\xi^2}\kappa_2 \left[ i |\xi_1|^2 \zeta_{i,0} + \sqrt{i} \xi_1^*\xi_2^* \zeta_{i-1,1} \right] 
\end{eqnarray}
and
\begin{eqnarray} \label{pop1}
	\dot{\zeta}_{i,1} &=& -\frac{\rm i}{2} \Big[ \sqrt{i+1} \Omega_a \zeta_{i+1,1} + \sqrt{i} \Omega_a^*  \zeta_{i-1,1} + \sqrt{2} \Omega_b \zeta_{i,2}  \nonumber \\
	 && + \Omega_b^* \zeta_{i,0} \Big] - \frac{1}{2\xi^2}\kappa_1 \Big[ \left(|\xi_1|^2 + i |\xi_2|^2\right) \zeta_{i,1} \nonumber \\
	 && + \sqrt{i+1} \xi_1 \xi_2 \zeta_{i+1,0} + \sqrt{2i} \xi_1^*\xi_2^* \zeta_{i-1,2} \Big] \nonumber \\
	&&-\frac{1}{2\xi^2}\kappa_2 \Big[ \left(|\xi_2|^2 + i |\xi_1|^2\right) \zeta_{i,1} - \sqrt{i+1} \xi_1\xi_2 \zeta_{i+1,0} \nonumber \\
	&&- \sqrt{2i} \xi_1^*\xi_2^* \zeta_{i-1,2} \Big] -\frac{1}{2} \kappa_m \zeta_{i,1} \, .
\end{eqnarray}
In the parameter regime given by Eq.~(\ref{condi}), states with photons in the $c_b$ mode evolve on a much faster time scale than states with population only in the $c_a$ mode. Consequently, the coefficients $\zeta_{i,j}$ with $j>1$ can be eliminated adiabatically from the system dynamics. Doing so and setting the right hand side of Eq.~(\ref{pop1}) equal to zero, we find that
\begin{eqnarray} \label{zeta1}
	\zeta_{i,1} &=& - {1 \over \kappa_{\rm m}} \left[ {\rm i} \Omega_b^* \zeta_{i,0} - \sqrt{i+1} \, \frac{\xi_1\xi_2}{\xi^2} \Delta \kappa \zeta_{i+1,0} \right] 
\end{eqnarray}
with $\Delta \kappa$ defined as 
\begin{eqnarray} \label{Deltak}
	\Delta \kappa &\equiv & \kappa_1 - \kappa_2 \, .
\end{eqnarray}
Substituting Eq.~(\ref{zeta1}) into Eq.~(\ref{pop0}), we find that the effective conditional Hamiltonian of the two cavities is now given by
\begin{eqnarray}\label{H_eff}
	H_{\rm cond} &=& \frac{1}{2} \hbar \Omega_{\rm eff} \, c_a + {\rm H.c.} -\frac{\rm i}{2} \hbar \kappa_{\rm eff} \,  c_a^{\dagger}c_a \, .
\end{eqnarray}
Up to first order in $1/\kappa_{\rm m}$, the effective Rabi frequency $\Omega_{\rm eff} $ and the effective decay rate $\kappa_{\rm eff} $ of the $c_a$ mode are given by
\begin{eqnarray}\label{H_eff2}
	\Omega_{\rm eff} &\equiv & \Omega_a + \frac{\xi_1\xi_2 \Delta \kappa}{\xi^2 \kappa_m} \, \Omega_b\, , \nonumber \\
        \kappa_{\rm eff} &\equiv &
	{1 \over \xi^2}\left[ \kappa_1|\xi_2|^2 + \kappa_2|\xi_1|^2 - \frac{|\xi_1 \xi_2|^2 \Delta \kappa^2}{\xi^2 \kappa_m} \right] \, . ~~
\end{eqnarray}
The decay rate $\kappa_{\rm eff}$ lies always between $\kappa_1$ and $\kappa_2$. If both cavities couple in the same way to their individual reservoirs, i.e.~when $\xi_1 = \xi_2$ and $\kappa_1 = \kappa_2$, then we have $\Omega_{\rm eff} = \Omega_a$ and $\kappa_{\rm eff} = \kappa_1$. 

Eq.~(\ref{zeta1}) shows that any population in the $c_a$ mode always immediately causes a small amount of population in the $c_b$ mode. Taking this into account, the reset operators in Eq.~(\ref{R_common}) become 
\begin{eqnarray}\label{R_common2}
	R_1 &=& \sqrt{\kappa_1} \, \frac{\xi_2}{\xi} \, \left[ 1 - \frac{|\xi_1|^2 \Delta \kappa}{\xi^2 \kappa_m} \right] \, c_a \, , \nonumber \\
	R_2 &=& - \sqrt{\kappa_2} \, \frac{\xi_1}{\xi} \, \left[ 1 + \frac{|\xi_2|^2 \Delta \kappa}{\xi^2 \kappa_m} \right] \, c_a \, , \nonumber \\
	R_{\rm m} &=& - \sqrt{\kappa_m} \, \frac{\xi_1\xi_2 \Delta \kappa}{\xi^2 \kappa_{\rm m}} \, c_a \, .
\end{eqnarray}
Substituting these and Eq.~(\ref{H_eff}) into the master equation (\ref{master}) we find that it indeed simplifies to
the master equation of a single cavity. Analogous to Eq.~(\ref{master}) we now have 
\begin{eqnarray}\label{master3}
	\dot{\rho} &=& - {{\rm i} \over \hbar} \left[ \, H_{\rm cond} , \rho \, \right] + \kappa_{\rm eff} \, c_a \, \rho \, c_a^\dagger \, ,
\end{eqnarray}
while Eqs.~(\ref{H_eff}) and (\ref{R_common2}) are analogous to Eqs.~(\ref{H_cond}) and (\ref{Rsingle}). The only difference to Section \ref{single_cav} is that the single mode $c$ is now replaced by the non-local common cavity field mode $c_a$, while $\Omega$ and $\kappa$ are replaced by $\Omega_{\rm eff}$ and $\kappa_{\rm eff}$ in Eq.~(\ref{H_eff2}). The $c_b$ mode no longer participates in the system dynamics and remains to a very good approximation in its vacuum state. 

Finally, let us remark that one way of testing the single-mode behavior of the two fiber-coupled cavities is to measure their stationary state photon emission rate $I$. Since their master equation is effectively the same as in the single-cavity case, this rate is under ideal decoupling conditions, i.e.~in analogy to Eq.~(\ref{III}), given by
\begin{eqnarray}\label{I3}
	I &=& |\Omega_{\rm eff} |^2 / \kappa_{\rm eff} \, .
\end{eqnarray}
If the decay rates $\kappa_1$ and $\kappa_2$ and the Rabi frequencies $\Omega_1$ and $\Omega_2$ are known, then the only unknown parameters in the master equation are the relative phase between $\xi_1$ and $\xi_2$, the ratio $|\xi_1/\xi_2|$, and the spontaneous decay rate $\kappa_{\rm m}$. These can, in principle be determined experimentally, by measuring $I$ for different values of $\Omega_1$ and $\Omega_2$ \footnote{The dependence of $I$ on the modulus squared of $\Omega_{\rm eff}$ means that it is not possible to measure the absolute values of $\xi_1$ and $\xi_2$ but this  is exactly as one would expect it to be. Also in the single optical cavity, the overall phase factor of the field mode is not known a priori and has in general no physical consequences.}.

\subsection{Effectiveness of the $c_b$ mode decoupling}

To conclude this section, we now have a closer look at how small $\kappa_{\rm m}$ can be with respect to the $\kappa_i$ and $\Omega_i$ whilst still decoupling the $c_b$ mode from the system dynamics. To have a criterion for how well the above described decoupling mechanism works we calculate in the following the relative amount of population in the $c_b$ mode when the laser-driven cavities have reached their stationary state with $\dot \rho = 0$. This means, we now consider the mean photon numbers  
\begin{eqnarray}
	n_a \equiv \langle c_a^{\dagger}c_a \rangle ~~{\rm and}~~ n_b \equiv \langle c_b^{\dagger}c_b \rangle 
\end{eqnarray}
and use the master equation to obtain rate equations which predict their time evolution. In order to obtain a closed set of differential equations, we need to consider the expectation values
\begin{eqnarray}
	k_a &\equiv& \frac{\rm i}{|\Omega_a|}\langle \Omega_a c_a - \Omega_a^* c_a^{\dagger} \rangle \, , \nonumber \\
	k_b &\equiv& \frac{\rm i}{|\Omega_b|}\langle \Omega_b c_b - \Omega_b^* c_b^{\dagger} \rangle \, , \nonumber \\
	m &\equiv& \frac{1}{\xi^2} \langle \xi_1\xi_2c_b^{\dagger}c_a + \xi_1^*\xi_2^*c_a^{\dagger}c_b \rangle \, , \nonumber \\
	l_a &\equiv& \frac{\rm i}{|\Omega_b|\xi^2}\langle \xi_1\xi_2\Omega_b c_a - \xi_1^*\xi_2^*\Omega_b^*c_a^{\dagger} \rangle \, , \nonumber \\
	l_b &\equiv& \frac{\rm i}{|\Omega_a|\xi^2}\langle \xi_1\xi_2\Omega_a c_b - \xi_1^*\xi_2^*\Omega_a^*c_b^{\dagger} \rangle 
\end{eqnarray}
in addition to $n_a$ and $n_b$. Doing so and using again Eqs.~(\ref{master2}), (\ref{H_cond-comm}), and (\ref{R_common}), we find that 
\begin{eqnarray}\label{rates}
	\dot{n_a} &=& \frac{|\Omega_a|}{2}k_a - \frac{1}{2}\Delta \kappa \, m - \kappa_an_a \, , \nonumber \\
	\dot{n_b} &=& \frac{|\Omega_b|}{2}k_b - \frac{1}{2}\Delta \kappa \, m - (\kappa_b + \kappa_m)n_b \, , \nonumber \\	
	\dot{k_a} &=& |\Omega_a| - \frac{1}{2}\Delta \kappa \, l_b - \frac{1}{2}\kappa_ak_a  \, , \nonumber \\
	\dot{k_b} &=& |\Omega_b| - \frac{1}{2}\Delta \kappa \, l_a - \frac{1}{2}(\kappa_b + \kappa_m)k_b \, , \nonumber \\
	\dot{m} &=& \frac{|\Omega_b|}{2} l_a + \frac{|\Omega_a|}{2} l_b -\frac{|\xi_1 \xi_2|^2}{\xi^4}\Delta \kappa \, [n_a+n_b] \nonumber \\
	&& - \frac{1}{2}(\kappa_1+\kappa_2+\kappa_m)m \, , \nonumber \\
	\dot{l_a} &=& \frac{1}{2\xi^2|\Omega_a|}[\xi_1\xi_2\Omega_b\Omega_a^* + \xi_1^*\xi_2^*\Omega_b^*\Omega_a] - \frac{|\xi_1 \xi_2|^2}{2\xi^4}\Delta \kappa \, k_b \nonumber \\	
	&& - \frac{1}{2}\kappa_al_a \, , \nonumber \\
	\dot{l_b} &=& \frac{1}{2\xi^2|\Omega_b|}[\xi_1\xi_2\Omega_b\Omega_a^* + \xi_1^*\xi_2^*\Omega_b^*\Omega_a] -\frac{|\xi_1 \xi_2|^2}{2\xi^4}\Delta \kappa \, k_a \nonumber \\
	&& - \frac{1}{2}\kappa_bl_b \, ,
\end{eqnarray}
where
\begin{eqnarray}
	\kappa_a &\equiv& \frac{1}{\xi^2}(\kappa_1|\xi_2|^2 + \kappa_2|\xi_1|^2) \, , \nonumber \\
	\kappa_b &\equiv& \frac{1}{\xi^2}(\kappa_1|\xi_1|^2 + \kappa_2|\xi_2|^2) 
\end{eqnarray}	
are the spontaneous decay rates of the $c_a$ and the $c_b$ mode, respectively. 

The stationary state of the system can be found by setting the right hand sides of the above rate equations equal to zero. However, the analytic solution of these equations is complicated and not very instructive. We therefore restrict ourselves in the following to the case where both cavities are driven by laser fields with identical Rabi frequencies and where both couple identically to the environment, i.e.
\begin{eqnarray}\label{Omega}
\Omega = \Omega_1 = \Omega_2  ~~ {\rm and} ~~\xi = |\xi_1| = |\xi_2| \, .
\end{eqnarray}
The remaining free parameters are a phase factor $\Phi$ between $\xi_1$ and $\xi_2$ defined by the equation
\begin{eqnarray} \label{phi}
\xi_2 = {\rm e}^{{\rm i} \Phi} \, \xi_1 
\end{eqnarray}
and the cavity decay rates $\kappa_1$, $\kappa_2$, and $\kappa_{\rm m}$. The reason that we restrict ourselves here to the case where the relative phase between the Rabi frequencies $\Omega_1$ and $\Omega_2$ equals zero, is that varying this phase has the same effect as varying the angle $\Phi$. 

\subsubsection{Identical decay rates $\kappa_1$ and $\kappa_2$}

\begin{figure}[t]
\begin{minipage}{\columnwidth}
\begin{center}
\resizebox{\columnwidth}{!}{\rotatebox{0}{\includegraphics{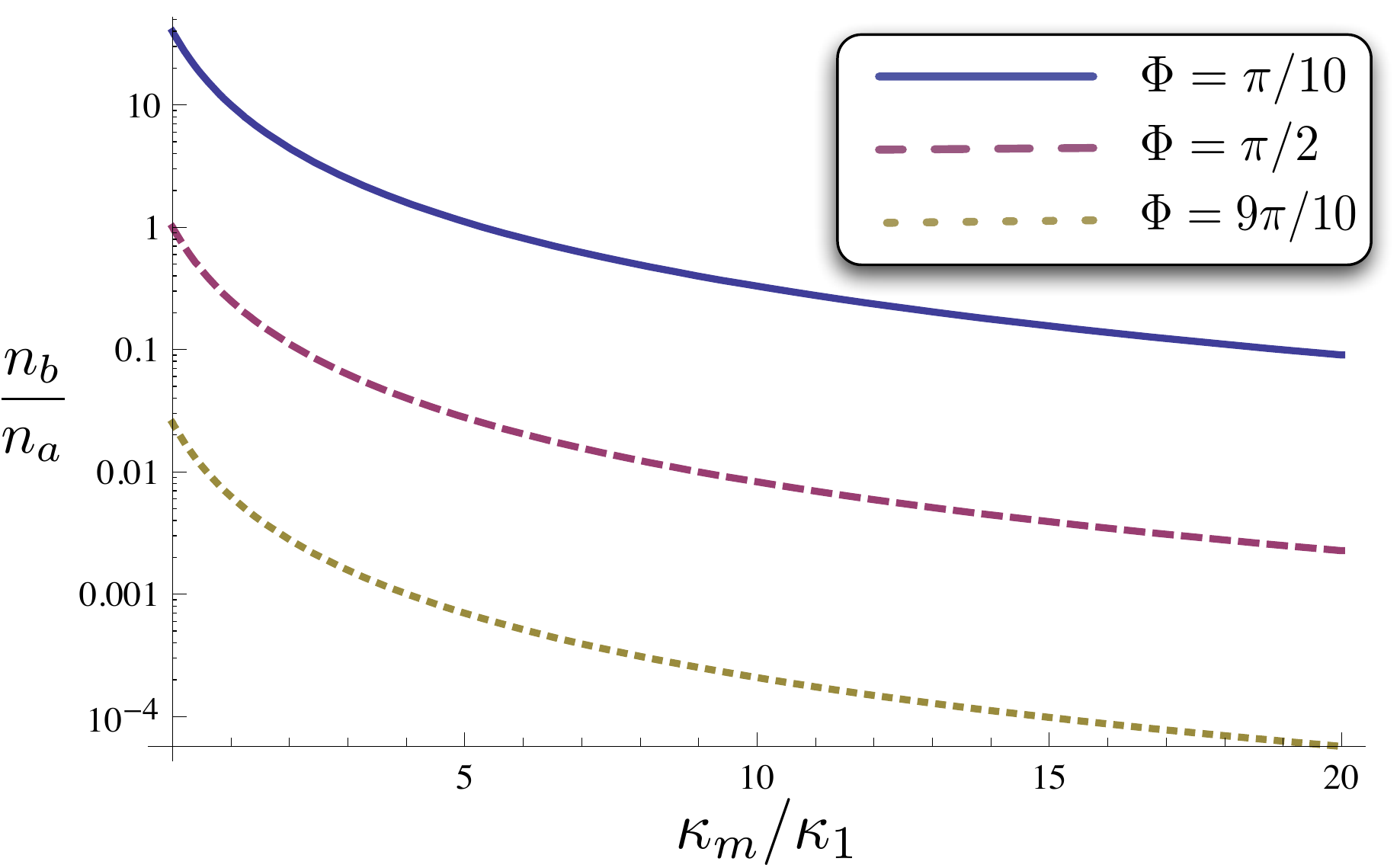}}}
\end{center}
\vspace*{-0.5cm} \caption{(Color online) Stationary state value of $n_b/n_a$ as a function of $\kappa_{\rm m}$ for $\xi_1 = \xi_2$ and $\Omega_1 = \Omega_2 = \kappa_1 = \kappa_2$ for three different values of $\Phi$ obtained from Eq.~(\ref{nanb1}).} \label{simplenb}
\end{minipage}
\end{figure}

To illustrate how these free parameters affect the robustness of the $c_b$ mode decoupling, we now analyse some specific choices of parameters. The first and simplest choice of parameters is to set the decay rates for both cavities the same. As in Eq.~(\ref{44}) we define
\begin{eqnarray}
\kappa \equiv \kappa_1 = \kappa_2
\end{eqnarray}
which implies $\Delta \kappa = 0$ and $\kappa_a = \kappa_b = \kappa$. Moreover, the rate equations in Eq.~(\ref{rates}) simplify to the four coupled equations
\begin{eqnarray}
	\dot{n_a} &=& \frac{|\Omega|}{\sqrt{2}}(1-\cos{\Phi})^{\frac{1}{2}}k_a - \kappa n_a \, , \nonumber \\
	\dot{n_b} &=& \frac{|\Omega|}{\sqrt{2}}(1+\cos{\Phi})^{\frac{1}{2}}k_b - (\kappa + \kappa_m)n_b \, , \nonumber \\
	\dot{k_a} &=& \sqrt{2}|\Omega|(1-\cos{\Phi})^{\frac{1}{2}} - \frac{1}{2}\kappa k_a \, , \nonumber \\
	\dot{k_b} &=& \sqrt{2}|\Omega|(1+\cos{\Phi})^{\frac{1}{2}} - \frac{1}{2}(\kappa + \kappa_m)k_b \, .
\end{eqnarray}
The stationary state of these equations can be calculated by setting these derivatives equal to zero. Doing so, we find that the mean number of photons in the $c_a$ and in the $c_b$ mode approach the values
\begin{eqnarray} \label{nanb1}
	n_a &=& (1-\cos{\Phi}) \, \frac{\Omega^2}{\kappa^2} \, , \nonumber \\
	n_b &=& (1+\cos{\Phi}) \, \frac{\Omega^2}{(\kappa + \kappa_{\rm m})^2} 
\end{eqnarray}
after a certain transition time. A measure for the effectiveness of the decoupling of the $c_b$ mode is given by the final ratio $n_b / n_a$ which is given by 
\begin{eqnarray}
	\frac{n_b}{n_a} &=& \frac{1+\cos{\Phi}}{1-\cos{\Phi}} \cdot \frac{\kappa^2}{(\kappa + \kappa_{\rm m})^2} \, .
\end{eqnarray} 
In general, this ratio tends to zero when $\kappa_{\rm m}$ becomes much larger than $\kappa$. There is only one exceptional case, namely the case where $\cos{\Phi} = 1$. This case corresponds to sole driving of the $c_b$ mode, where the stationary state of the $c_a$ mode corresponds to $n_a = 0$.

\begin{figure}[t]
\begin{minipage}{\columnwidth}
\begin{center}
\resizebox{\columnwidth}{!}{\rotatebox{0}{\includegraphics{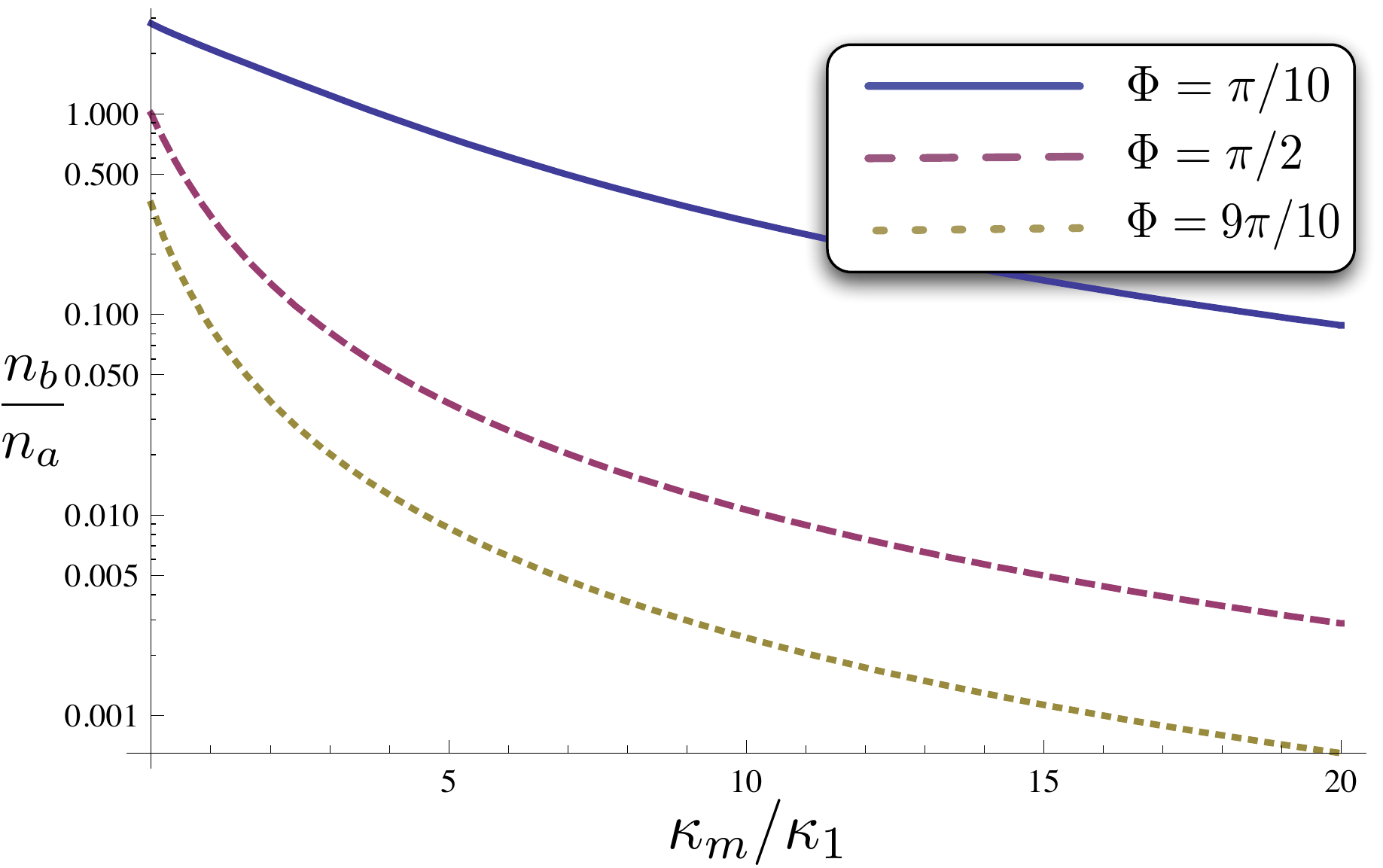}}}
\end{center}
\vspace*{-0.5cm} \caption{(Color online) Stationary state value of $n_b/n_a$ as a function of $\kappa_{\rm m}$ for $\xi_1 = \xi_2$, $\Omega_1 = \Omega_2 = \kappa_1$, and $\kappa_2 = 0.5 \, \kappa_1$. As in Fig.~\ref{simplenb}, we observe a very rapid drop of the relative population in the $c_b$ mode as $\kappa_{\rm m} $ increases.} \label{nbk2-05}
\end{minipage}
\end{figure}

This behavior is confirmed by Fig.~\ref{simplenb} which shows the steady state value of $n_b/n_a$ in Eq.~(\ref{nanb1}) as a function of $\kappa_{\rm m}$ for three different values of $\Phi$. In all three cases, the mean photon number in the $c_b$ mode decreases rapidly as $\kappa_{\rm m}$ increases. This is an indication of the robustness of the decoupling of the $c_b$ mode. It shows that this decoupling does not require an alignment of the driving lasers. However, as already mentioned above, one should avoid sole driving of the $c_b$ mode. Indeed we find relatively large values for $n_b/n_a$ when the angle $\Phi$ is relatively small. The case $\Phi = \pi/2$ corresponds to equal driving of both common modes. In this case we have $n_b/n_a < 0.01$ when $\kappa_{\rm m}$ is at least eight times larger than $\kappa$ which is a relatively modest decoupling condition. Close to the perfect alignment case (with $\Phi = \pi$) which we discussed in detail in the previous subsection, $n_b / n_a$ is even smaller than in the other two cases. For $\Phi = 0.9 \, \pi$ and $\kappa_{\rm m} > 8 \, \kappa$, we now already get $n_b/n_a \ll 0.001$.

\subsubsection{Different decay rates $\kappa_1$ and $\kappa_2$}

\begin{figure}[t]
\begin{minipage}{\columnwidth}
\begin{center}
\resizebox{\columnwidth}{!}{\rotatebox{0}{\includegraphics{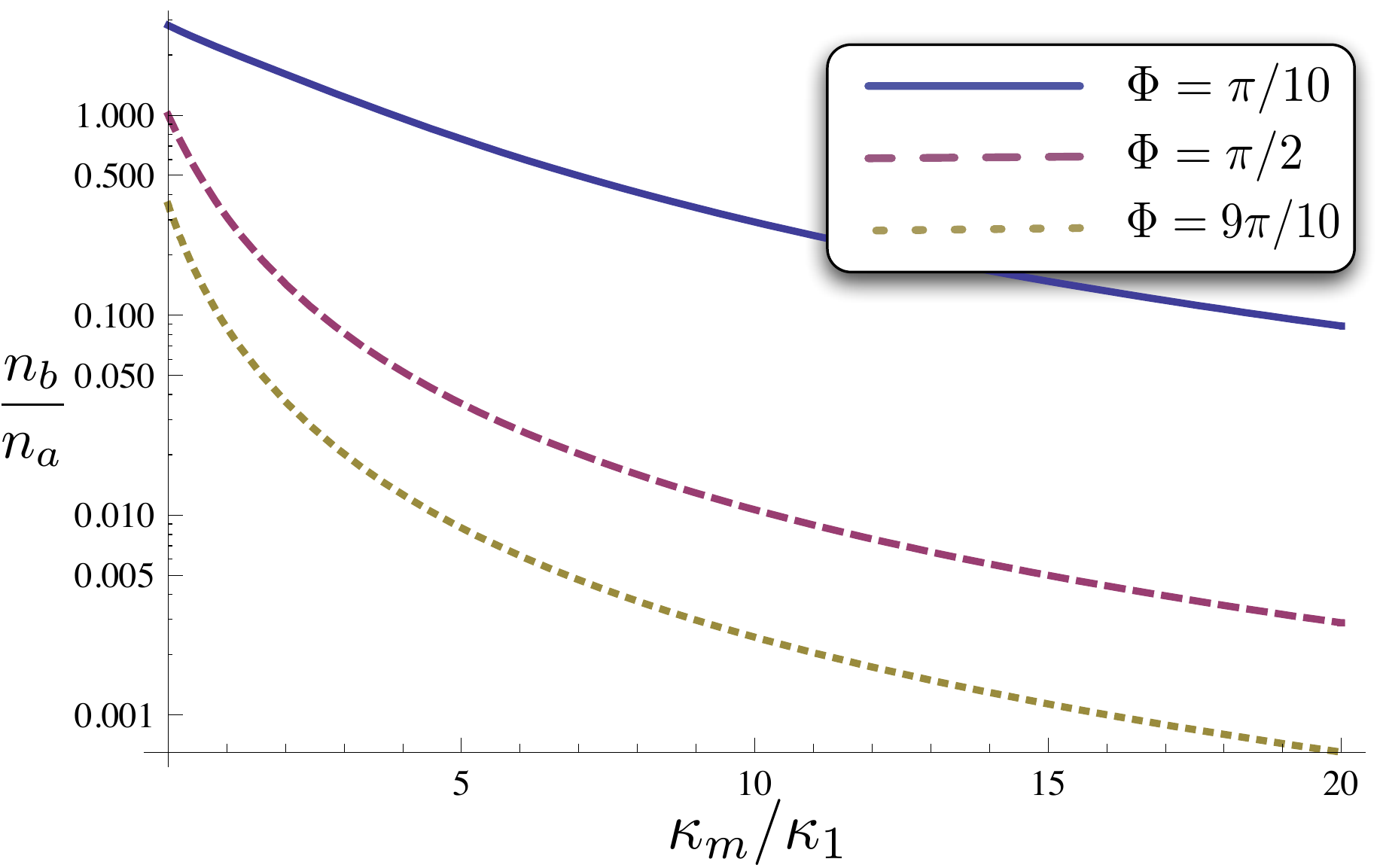}}}
\end{center}
\vspace*{-0.5cm} \caption{(Color online) Stationary state value of $n_b/n_a$ as a function of $\kappa_{\rm m}$ for $\xi_1 = \xi_2$, $\Omega_1 = \Omega_2 = \kappa_1$, and $\kappa_2 = 1.5 \, \kappa_1$. As in Figs.~\ref{simplenb} and \ref{nbk2-05}, $n_b/n_a$ decreases rapidly as $\kappa_{\rm m} $ increases. The main difference to Fig.~\ref{nbk2-05} is that we now have $\Delta \kappa < 0$ instead of having $\Delta \kappa > 0$.} \label{nbk2-15}
\end{minipage}
\end{figure}

In the above case with $\Delta \kappa = 0$, there is no transfer of photons between the two modes. To show that this is not an explicit requirement for the decoupling of the $c_b$ mode, we now have a closer look at the case where $\Delta \kappa \neq 0$ and where mixing between both common cavity modes occurs. Let us first have a look at the case where $\Phi = 0$ and where only the $c_b$ mode is driven. In this case, we expect $\Delta \kappa$ to result in an enhancement of the single mode behavior compared to the $\Delta \kappa = 0$ case. The reason is that the effective Rabi frequency $\Omega_{\rm eff}$ in Eq.~(\ref{H_eff2}) is now always larger than zero such that $n_a$ no longer tends to zero when $\Phi \to 0$. Different from this, we expect the stationary state value of $n_b/n_a$ to increase when $\Phi = \pi$. The reason for this is that this case now no longer corresponds to perfect alignment which required $\Delta \kappa = 0$ (c.f.~Eq.~(\ref{44})). This behavior of the two fiber-coupled cavities is confirmed by Figs.~\ref{nbk2-05} and \ref{nbk2-15} which have been obtained by setting the time derivatives of the original rate equations (\ref{rates}) equal to zero. For the parameters considered here, the introduction of $\Delta \kappa$ has no effect on the effectiveness of the decoupling of the $c_b$ mode when $\Phi = \pi/2$ and both modes are equally driven by laser fields. 

\section{Conclusions} \label{conc}

In conclusion, we have presented a scheme that couples two cavities with a single-mode fiber coated with two-level atoms (c.f.~Fig.~\ref{scheme}) or a waveguide (c.f.~Fig.~\ref{scheme2}). Since there are two cavities, the description of the system requires two orthogonal cavity field modes. These could be the individual cavity modes with the annihilation operators $c_1$ and $c_2$ or common modes with the annihilation operators $c_a$ and $c_b$ in Eq.~(\ref{com_modes}). Here we consider the case where the connection between the cavities constitutes a reservoir for only one common cavity field mode but not for both. If this mode is the $c_b$ mode, it can have a much larger spontaneous decay rate than the $c_a$ mode which does not see this reservoir. A non-local resonator is created, when operating the system in the parameter regime given by Eq.~(\ref{condi}), where the $c_b$ mode can be adiabatically eliminated from the system dynamics, thereby leaving behind only the $c_a$ mode. 

The purpose of the atoms which coat the fiber is similar to their purpose in Ref.~\cite{Franson}, namely to measure its evanescent electric field destructively, although here there is no need to distinguish between one and two photon states. These measurements should occur on a time scale which is long compared to the time it takes a photon to travel from one resonator to the other. One can easily check that this condition combined with Eq.~(\ref{condi}) poses the following upper bound on the possible length $R$ of the fiber: 
\begin{eqnarray} 
{R \over c} \, \ll \, {1 \over \kappa_{\rm m}} \, \ll \,  {1 \over \kappa_1} , \, {1 \over \kappa_2} \, .
\end{eqnarray}
Here $\kappa_1$, $\kappa_2$, and  $\kappa_{\rm m}$ are the spontaneous cavity decay rates through the outcoupling mirrors of cavity 1 and cavity 2 and through the fiber reservoir, respectively, while $c$ denotes the speed of light. This means, the possible length $R$ of the fiber depends on how good the cavities are. For good cavities, $R$ could be of the order of several meters. However, the upper bound for $R$ depends also on the fiber diameter and the quality of the mirrors. The reason is that the fiber should not support a too wide range of optical frequencies, i.e.~the fiber should support only one standing wave with frequency $\omega_{\rm cav}$ and not two degenerate ones.

There are different ways of seeing how the coated fiber removes one common cavity field mode from the system dynamics. One way is to compare the setup in Fig.~\ref{scheme} with the two-atom double-slit experiment by Eichmann {\em et al.} \cite{Eichmann} which has been analysed in detail for example in Refs.~\cite{Schoen,pachos00}. In this experiment, two atoms are simultaneously (i.e.~in phase) driven by a single laser field and emit photons into different spatial directions. The emitted photons are collected on a photographic plate which shows intensity maxima as well as completely dark spots. A dark spot corresponds to a direction of light emission where the atomic excitation does not couple to the free radiation field between the atoms and the screen due to destructive interference. The setup in Fig.~\ref{scheme} creates an analogous situation: the photons inside the two resonators are the sources for the emitted light, thereby replacing the atomic excitation. Moreover, the light inside the fiber is equivalent to a single-mode (i.e.~one wave vector ${\bf k}$) of the free radiation field in the double slit experiment. There is hence one common resonator mode -- the $c_b$ mode -- which does not couple to the fiber. 

This paper describes the setups in Figs.~\ref{scheme} and \ref{scheme2} in a more formal way. Starting from the Hamiltonian as in Ref.~\cite{Pellizzari} for the cavity-fiber coupling but considering the radiation field inside the fiber as a reservoir we derive the master equation for the time evolution of the photons in the optical cavities. After the adiabatic elimination of one common cavity mode, namely the $c_b$ mode, due to overdamping of its population, we arrive at a master equation which is equivalent to the master equation of a single laser-driven optical cavity.

A concrete measure for the quality of the decoupling of the $c_b$ mode is the stationary state value of $n_b/n_a$, where $n_a$ and $n_b$ are the mean numbers of photons in the $c_a$ and the $c_b$ mode, respectively, when both cavities are driven by a resonant external laser field. Our calculations show that this ratio can be reduced significantly by a careful alignment of the driving lasers. However, even when both cavity modes couple equally to two external laser fields, $n_b/n_a$ can be as small as $0.01$ even when $\kappa_{\rm m}$ is only one order of magnitude larger than $\kappa_1$, $\kappa_2$, and the Rabi frequencies of the driving lasers. This parameter regime consequently does not require any alignment and is very robust against parameter fluctuations.

Possible applications of this setup become apparent when one places for example atomic qubits, single quantum dots, or NV color centers into each cavity. These would feel only a common cavity field mode and interact as if they were sitting in the same resonator. Such a scenario has applications in quantum information processing, since it allows to apply quantum computing schemes like the ones proposed in Refs.~\cite{Metz,Metz2} which would otherwise require the shuttling of qubits in and out of an optical resonator to spatially separated qubits. 

In recent years, a lot of progress has been made in the laboratory and several atom-cavity experiments which operate in the strong coupling regime have already been realised \cite{Rempe,Kimble,Chapman,Trupke,Reichel,Meschede,Blatt}. Some of these experiments have become possible due to new cavity technologies. Optical cavities with a very small mode volume are now almost routinely mounted on atom chips using novel etching techniques and specially coated fiber tips \cite{Trupke,Reichel}. These can in principle also be coupled to miniaturised ion traps \cite{Schmidt-Kaler} or telecommunication-wavelength solid-state memories \cite{Gisin}. Alternatively, strong couplings are achieved in the microwave regime with so-called stripline cavities \cite{Schoelkopf}. In several of these experiments, the coupling of cavities via optical fibers or waveguides as illustrated in Fig.~\ref{scheme} could be a possible next step. \\[0.5cm]

{\em Acknowledgement.} We thank W.~Altm\"uller, A.~Kuhn, and M. Trupke for very helpful and stimulating discussions. J. B. acknowledges financial support from the European Commission of the European Union under the FP7 STREP Project HIP (Hybrid Information Processing). A. B. acknowledges a James Ellis University Research Fellowship from the Royal Society and the GCHQ. Moreover, this work was supported in part by the European Union Research and Training Network EMALI.

\end{document}